\documentclass{icrc2009}
\usepackage{graphicx}   
\usepackage{caption}    
\usepackage[font=footnotesize]{subfig} 
\usepackage{fixltx2e}
\usepackage{url}

\newcommand{\shorttitle}[1]%
{\markboth{Proceedings of the 31\MakeLowercase{$^{st}$} ICRC, {\L}\'{o}d\'{z} 2009}{#1} }
\newcommand{\etal}{\MakeLowercase{\textit{et al. }}} 

\begin{document}
\title{The time-space structure of pulses in Cherenkov light detectors}

\author{\IEEEauthorblockN{D.A. Podgrudkov\IEEEauthorrefmark{1},
                          L.G. Dedenko\IEEEauthorrefmark{1},
                          T.M. Roganova\IEEEauthorrefmark{1} and
                          G.F. Fedorova\IEEEauthorrefmark{1}}
                                                      \\
\IEEEauthorblockA{\IEEEauthorrefmark{1}D.V. Skobeltsyn Institute of Nuclear Physics, MSU,\\ Leninskie Gory,
119992 Moscow, Russia}}

\shorttitle{D.A. Podgrudkov \etal The time-space structure...}
\maketitle

\begin{abstract}

Here the results of calculations of pulses in Cherenkov light detectors for the
Yakutsk array are presented. As long as  the Vavilov-Cherenkov light is used to
calibrate signals in scintillation detectors at the Yakutsk array it is vital for
these measurements to be precise. The validity of measurements of the signal
$Q(400)$ used as the estimation parameter at the Yakutsk array has been confirmed.
Our calculations show that the width of time pulses increases from nearly 50~ns at
 a distance of 100~m from the shower axis up to 700~ns at 1000~m.

\end{abstract}


\section{Introduction}

It is a known problem of calibration of energy estimations of extensive air
showers (EAS). At the Yakutsk array the Vavilov-Cherenkov light is used to
calibrate signals in scintillation detectors \cite{1}. Although our calculations
show that for the same flux of Vavilov-Cherenkov light there should be much
higher signals in scintillation detectors than it is measured in the experiment \cite{2}.

First, this contradiction may have different explanations. The hadron interaction
model may not correctly describe the real processes at high energies. This
question will be probed at LHC once it is operational. Alternatively, we may
have a different chemical composition of primary cosmic rays -- all our
calculations were undertaken in the assumption of light chemical composition
at high energies. Although recent experiments claim that the primary
particles at those energies are protons \cite{3,4} (with exclusion of \cite{5}).

 \begin{figure}[!t]
  \centering
  \includegraphics[width=8cm]{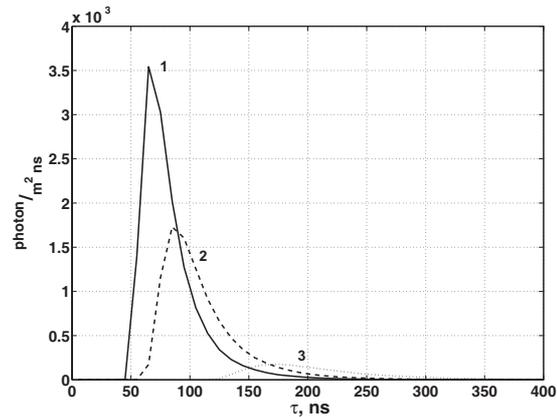}
  \caption{Pulses of Vavilov-Cherenkov light at 400~m distance from the shower axis.
Showers induced by primary gamma with energy 10~GeV starting from different
depth (solid curve 1 -- 350~g/cm$^2$, dashed curve 2 -- 550 g/cm$^2$, dotted curve 3 -- 750 g/cm$^2$)}
  \label{Fig. 1}
 \end{figure}

\begin{figure}[!t]
  \centering
  \includegraphics[width=8cm]{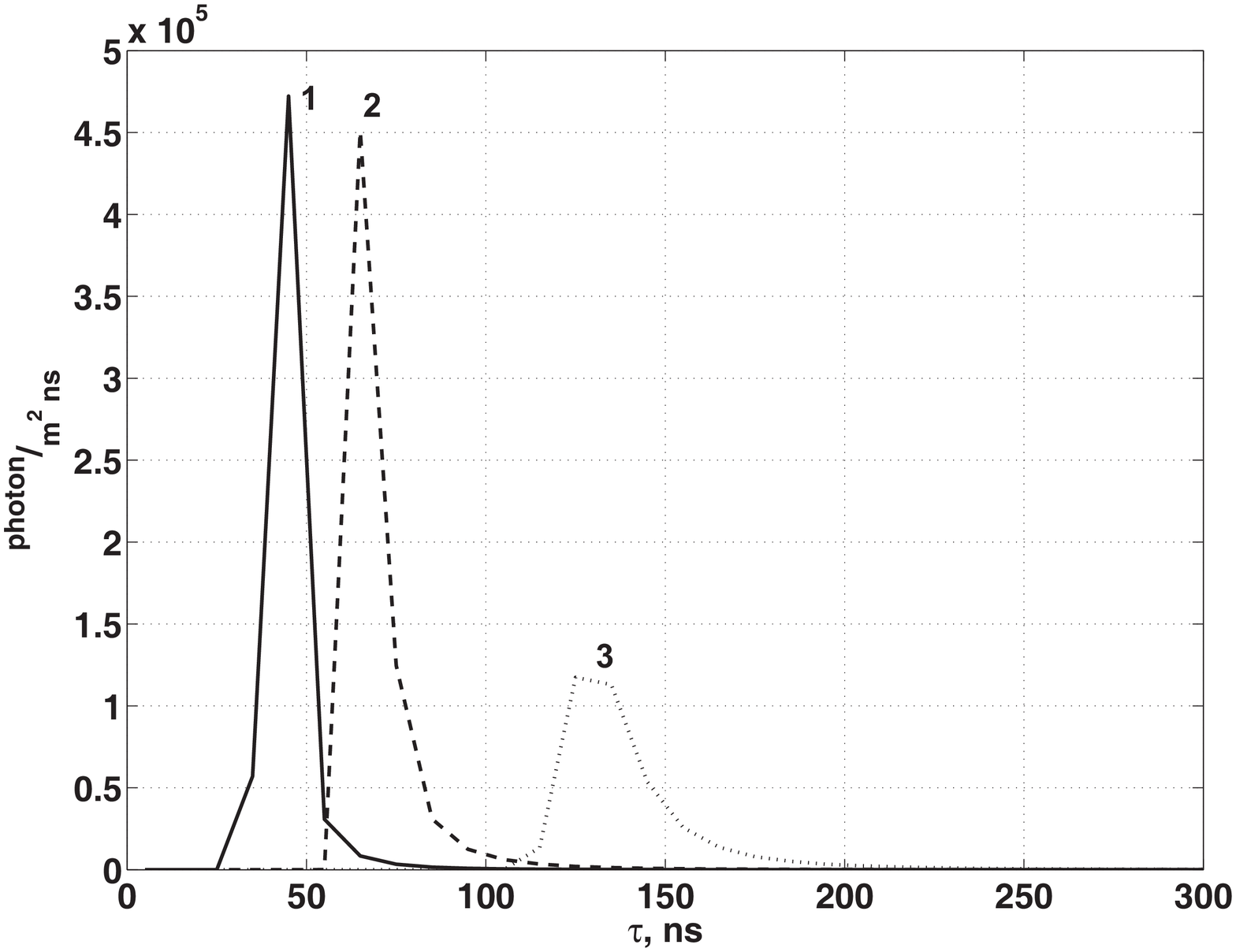}
  \caption{Pulses of Vavilov-Cherenkov light at 400~m distance from the shower axis.
Showers induced by primary electron with energy 316~MeV starting from different
depth (solid curve 1 -- 350~g/cm$^2$, dashed curve 2 -- 550 g/cm$^2$, dotted curve 3 -- 750 g/cm$^2$)}
  \label{Fig. 2}
 \end{figure}

\begin{figure}[!t]
  \centering
  \includegraphics[width=8cm]{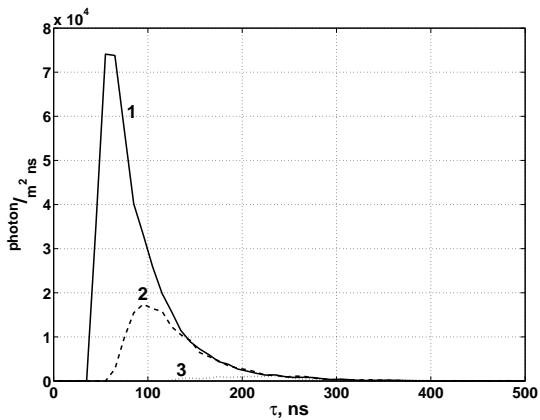}
  \caption{Pulses of Vavilov-Cherenkov light at 400~m distance from the shower axis.
Showers induced by primary muon with energy 10~GeV starting from different
depth (solid curve 1 -- 350~g/cm$^2$, dashed curve 2 -- 550 g/cm$^2$, dotted curve 3 -- 750 g/cm$^2$)}
  \label{Fig. 3}
 \end{figure}

Each detector array encounters the problem of full signal registration. Each
detector after being triggered by some event (a particle hits the detector,
the counting rate exceeds some level, a signal comes from another detector,
etc.) collects the signal for some period of time, a so-called time gate.
Time gates should be wide enough to collect possibly all particles coming to
the detector. On the other hand, since we always have a background from low
energy cosmic rays or the local sources (radioactivity, light pollution,
etc.), the time gates should not be too wide in order to keep the signal to
noise ratio high enough. So the time gates should be of about the disk
thickness. At most detector arrays such as Haverah Park, Volcano Ranch,
Yakutsk, the time gates were set to about 2 $\mu $s.

A. Watson in \cite{6} suggested that the time gates at the Yakutsk array are too
narrow to detect all particles and thus the signals are underestimated, that
leads to an overestimation of energy of the primary particles. A rapid change in
steepness of the measured lateral distribution functions of signals and the fact
that some of the showers detected at Haverah Park had signal width more than 2.2~$\mu $s \cite{6}
are claimed to support this suggestion.

In our previous paper \cite{7} we have studied the problem of form and width of
pulses in scintillation detector. In a few words though the signals in
scintillation detectors at large distances from the shower axis (more than 1~
km) exceed 2~$\mu $s (at 600~m signal width is well within 2~$\mu $s) and only
part of the signal is measured the difference between the theory and the
experiment is still too high to be explained solely by partial signal
measurements. Therefore, additional analysis was required. In this paper we
investigated another question -- the form of the Cherenkov light pulse at the
detector.

\section{Method of simulations}

As well as in the previous work in this study we used CORSIKA 6.500 \cite{8} with
QGSJET-II \cite{9}--\cite{13} for high energy and Gheisha-2002d \cite{14} for low
energy calculations. Parameters of the atmosphere and the magnetic field were set
to fit the conditions of the Yakutsk array. The time delay of arrival of the
Vavilov-Cherenkov photons was calculated as the time between its arrival at
the observation level and the time of arrival of the very first particle at
the observation level. All calculations were carried out in the framework of
a multilevel hybrid scheme \cite{15}, i.e. we calculated the database of Vavilov-Cherenkov
light pulses at different distances from the shower axis from low energy showers
from primary electrons, gamma and muons (some examples are shown in Fig. 1--3)
and using source functions (also calculated in our previous simulations \cite{7}) we
calculated the impulses for high energy showers.

Source function describes the number of particles with energy in interval $[E ; E+dE]$
born in the atmosphere at depth $[x ; x+dx]$. The database of the Vavilov-Cherenkov
light pulses $C_{e,\gamma,\mu}(\tau,r,E,x)$ describes the Vavilov-Cherenkov photon
flux from a shower induced at depth $x$ by electron or gamma or muon with the energy $E$
at the distance from the shower axis $r$ at moment of time $\tau$. Time starts with
the arrival of the very first particle at the observation level. Having the analytical
source functions of low energy particles from ultra high energy shower and databases
of the Vavilov-Cherenkov light pulses from low energy showers one can calculate pulses
from ultra high energy shower:

$F(r,\tau)=\int\limits_{0}^{x_0}dx\int\limits_{E_{min}}^{E_{thr}}S_{e}(E,x)C_{e}(\tau,r,E,x)+\ \ \ \ \ \ \ \ \ \ \ \ \ $

$\ \ \ \ \ \ \ \ \ \ +\int\limits_{0}^{x_0}dx\int\limits_{E_{min}}^{E_{thr}}S_{\gamma}(E,x)C_{\gamma}(\tau,r,E,x)+$

$\ \ \ \ \ \ \ \ \ \ +\int\limits_{0}^{x_0}dx\int\limits_{E_{min}}^{E_{thr}}S_{\mu}(E,x)C_{\mu}(\tau,r,E,x).$

Here $E_{thr}$ - some reasonable maximum energy threshold, maximum energy of particles in source functions; $E_{min}$ -
some low energy threshold (naturally in case of the Vavilov-Cherenkov light for electrons and gammas it is 21 MeV, for muons - 4 GeV);
$S_{e,\gamma,\mu}(E,x))$ - source functions of electrons, gammas and muons respectively. But if the source functions were obtained
numerically as a table of parameters (energy $E_{i}$, depth of generation $x_{i}$ and weight $w_{i}$ if thinning procedure
was applied) of low energy particles generated in the shower than this integral converts to a sum:

$F(r,\tau)=\sum\limits_{i} w_{i}C_{e}(\tau,E_{i},x_{i},r)+\ \ \ \ \ \ \ \ \ \ \ $

$\ \ \ \ \ \ \ \ \ \ +\sum\limits_{j} w_{j}C_{\gamma}(\tau,E_{j},x_{j},r)+\ \ \ $

$\ \ \ \ \ \ \ \ \ \ +\sum\limits_{k} w_{k}C_{\mu}(\tau,E_{k},x_{k},r).\ \ \ $

After all theese calculations for vertical showers from primary protons we obtained the following results.

\section{Results }
\begin{figure}[!t]
  \centering
  \includegraphics[width=8cm]{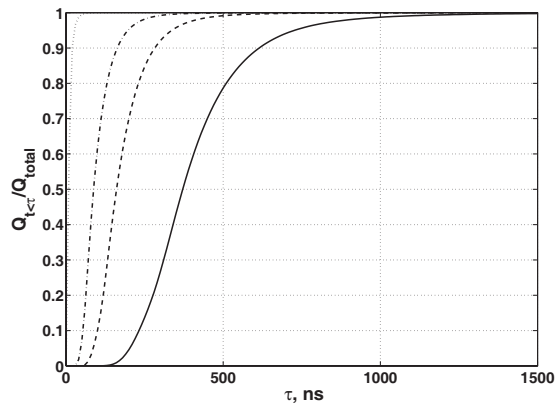}
  \caption{Fraction of Vavilov-Cherenkov light collected within time $\tau $ at the
different distances from shower axis (doted curve -- 100 m, dash-dotted
curve -- 400 m, dashed curve -- 600 m, solid curve -- 1000 m)}
  \label{Fig. 4}
 \end{figure}

\begin{figure}[!t]
  \centering
  \includegraphics[width=8cm]{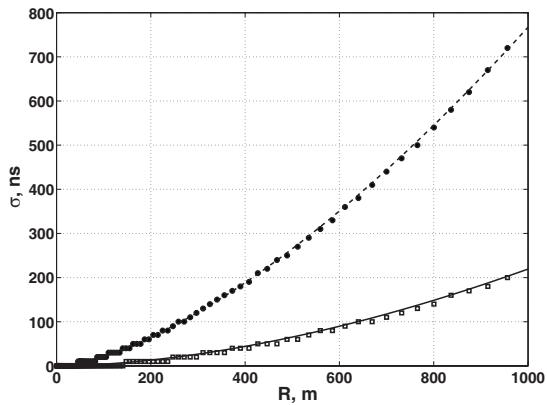}
  \caption{The Vavilov-Cherenkov light forefront (squares) and backfront (circles) and
their approximations}
  \label{Fig. 5}
\end{figure}

In Fig. 4 the fraction of Vavilov-Cherenkov light collected within time $\tau $
at different distances from the shower axis is shown. As one can see the width
of the Vavilov-Cherenkov light impulse at the distance of 100~m from the shower
axis is less than 50~ns, at 400~m -- 200~ns, at 600~m -- 400~ ns and at 1000~m
it is less than 1~$\mu $s. Therefore, the time gates at the Yakutsk array
detectors are set wide enough to collect all of the Cherenkov photons from
EAS, thus in the experiment the Vavilov-Cherenkov light is measured correctly.

In addition we have calculated the form of the Vavilov-Cherenkov light forefront
and backfront (see Fig. 5). As the forefront arrival time we took the moment
when 5\% of total signal is collected and the backfront was considered as
the moment when 95\% of all Cherenkov photons were collected. We found that
the best fit is achieved by the power function $\sigma = a\cdot R^{b}$ (with
$a_{f} = 1.63\cdot 10^{-3} $, $b_{f} $= 1.71, $a_{b} = 3.95\cdot 10^{-2} $ and
$b_{b} $= 1.43 for forefront and backfront respectively) rather than with a
spherical front.

\section{Conclusion}

The width of the Vavilov-Cherenkov light pulse does not exceed 1~$\mu $s at the
distance of 1000~m from the shower axis. Therefore, at the Yakutsk array all the
\newpage
Cherenkov photons from EAS are collected and the time gates of the Cherenkov
detectors may be set even lower.

Cherenkov light forefront and backfront may be approximated by a power
function $\sigma = a\cdot R^{b}$ (with $a_{f} = 1.63\cdot 10^{-3} $, $b_{f} $= 1.71,
$a_{b} = 3.95\cdot 10^{-2}$ and $b_{b} $= 1.43 for forefront and backfront respectively) with a better accuracy than with a spherical front.

\section{Acknowledgements}

 Authors thank RFBR (grant 07-02-01212) and G.T. Zatsepin LSS
 (grant 959.2008.2) for support.\\

\end{document}